\documentclass[aps,prl,a4paper,twocolumn,superscriptaddress,showkeys,longbibliography]{revtex4-2}

\usepackage[colorlinks=true,bookmarks= false,citecolor=blue,linkcolor=red,hyperfootnotes=true,urlcolor=blue]{hyperref}
\usepackage{dcolumn}
\usepackage{bm}
\usepackage{physics}
\usepackage{graphicx}
\usepackage{dcolumn}
\usepackage{bm}
\usepackage{wasysym}
\usepackage{lipsum}
\usepackage{enumitem}
\usepackage{booktabs} 
\usepackage{gensymb}
\usepackage{xcolor}
\usepackage{amssymb}
\usepackage{xcolor}
\usepackage{amsmath}
\usepackage{subfigure}

\usepackage{mathtools}
\usepackage{soul}
\usepackage{float}

\usepackage{changes}

\definecolor{darkgreen}{RGB}{0,100,0}

\begin{document}

\preprint{APS/123-QED}

\title{Emergent scale and anomalous dynamics in certain quasi-periodic systems}

\date{\today}

\newcommand*\samethanks[1][\value{footnote}]{\footnotemark[#1]}

\author{Parvathy S Nair}
\affiliation{Department of Physics, Indian Institute of Science Education and Research (IISER)
Tirupati, Tirupati - 517507, Andhra Pradesh, India}
\author{Dintomon Joy}
\affiliation{Department of Physics, Indian Institute of Science Education and Research (IISER)
Tirupati, Tirupati - 517507, Andhra Pradesh, India}
\affiliation{Department of Physics, St. Thomas College Palai, Pala - 686574, Kerala, India}
\author{Sambuddha Sanyal}\thanks{Author to whom correspondence should be addressed} \email{sambuddha.sanyal@iisertirupati.ac.in}
\affiliation{Department of Physics, Indian Institute of Science Education and Research (IISER)
Tirupati, Tirupati - 517507, Andhra Pradesh, India}
\affiliation{Center for Atomic, Molecular and Optical Sciences $\&$ Technologies(CAMOST), Joint initiative of IIT Tirupati $\&$ IISER Tirupati, Yerpedu, 517619,
Andhra Pradesh, India}

\date{\today}
%
\begin{abstract}
We study localisation transition in a class of quasi-periodic systems that has
two competing periodic scales. We show that such class of systems show a re-entrant localisation transition where the energy scale of transition is set by the periodicities of these two scales. Furthermore we show dynamical properties in these systems, exhibits
various kinds critical dynamics including sub-diffusive, super-diffusive and diffusive spread of an initially localised wave-packet. Finally we show that these characteristics of quasi-periodic systems with two periodic scales can be realised
within the regime of current experiments.
\end{abstract}

\maketitle

\paragraph*{\emph{Introduction.}}
Quasi-periodic(QP) models are often seen as  
archetypal templates\cite{hobbyhorse} to study delocalisation-localisation transitions(DLT) in quantum many body systems with broken translation symmetry. Localisation was
 originally envisaged in lattice models\cite{Andersonloc} where translation symmetry is broken by random on-site energies. However, DLT in these model occurs only in dimension $d>2$ \cite{GO4}, where a finite amount of disorder localises some of the single particle states, separated from the delocalised states by a mobility edge; with increasing disorder all the states in the spectrum become localised. This disorder induced quantum phase transition is known as Anderson localization transition(ALT)\cite{Anderson_transitions}. In a paradigmatic study, Aubry and Andre showed that in a lattice model (AA model)\cite{aa1} with a quasi-periodic on-site potential of periodicity that is incommensurate to the lattice periodicity,  a similar DLT occurs even in $d=1$ when the amplitude of the potential is increased from low to high. The interest in such QP models has been escalating for the last two decades mainly for the following three reasons. First: the deterministic nature and lower dimensional realization of the localisation transition in QP models allows one to study such phenomena at a much cheaper computational cost compared to ALT. Second: the characteristics of DLT in QP models are very distinct from ALT apart from lower dimensional realization albeit with certain similarities\cite{aa_multifrac}. While ALT is an interference phenomenon, the DLT in QP models is a property of the solution of the Schrodinger equation\cite{aa_math_1,avila_global,ten_martini}. This feature in QP systems is particularly interesting as it leads to unique signatures in dynamics and transport\cite{sanyal_aa} and is known as a distinct universality class\cite{aa_multifractal_wf,Anderson_transitions,aa_multifractal_spectrum}\footnote{Moreover, the localisation properties of the many-body states of such QP models can be immune to various natural perturbations\cite{avalanche_chandran,avalanche_dassarma} that arises in real systems.}. Third: due to unprecedented progress in cold atom experiments in recent times\cite{aa_expt_bec,aa_expt_1,aa_expt_2d,aa_expt_slow_dynamics,
 aa_expt_spme,aa_expt_spme1,aa_expt_me,aa_expt_photonic,aa_expt_boson,
 aa_expt_subdiffusion_bosonic,aa_expt_cascade}, QP models can be engineered and controlled in real-world systems.

Generally, QP models are described by three types of parameters. First is an energy scale, the second is a frequency or periodicity, and the third is the phase that gives a uniform shift to the periodic/quasi-periodic modulation. In the original AA model\cite{aa1}, there is only one energy scale i.e. the magnitude of potential amplitude(in the scale of unit hopping amplitude), one frequency parameter given by an irrational number that characterizes the quasi-periodicity and one phase shift parameter. The onset of DLT in the AA model depends only on the potential amplitude and is even independent of energy i.e. there is no mobility edge(ME). Notably, the frequency, an easily controllable quantity in experiments, only affects the multifractality of the wave function at the critical point\cite{aa_multifrac,aa_multifrac_1}.
Over the last four decades, many QP models\cite{aa_me_slow,aa_me_bichrome_1,aa_nnn_1,aa_lr_1,aa_bichrome_economou,aa_multiharmonic_1,
aa_me_bichrome_2,aa_me_bichrome_3,aa_me_bichrome_4,gaa,aa_mosaic} have been introduced with more than one of each kind of parameters, while these QP models show a wide range of exotic phenomena what they have in common is the presence of a ME.  \emph{ Interestingly, from all the studies till now, it is known that the energy scale of DLT in these extended models doesn't depend on the frequency or phase parameter}. 

In this letter, we unearth the rich and novel characteristics of a class of extended AA models that has two energy scales that are modulated with two different frequencies and is strictly one-dimensional with only nearest-neighbour hopping. We study the statics and dynamics of four models from this class. The main results of our study are summarised as follows: (1) We demonstrate that the frequency scale of these models surprisingly controls the energy scale of the ME. (2) We show that the competition between two periodic energy scales can give rise to re-entrant localisation phenomena in the models of this class. While such re-entrant phenomena were discovered recently in two examples of this class of models\cite{Goblot2020,PhysRevLett.126.106803,2109.09621}, we extend this list and also show that the frequency can not only control the energy scale of the onset of ME but also control the number of re-entrant transitions. (3) We find a wide range of dynamical behaviour in these models that include diffusive, sub-diffusive, and super-diffusive transport, and also a different regime of anomalous transport at early and late time dynamics. We further propose rich dynamical phase diagrams for these four models.

\paragraph*{}
In the rest of this letter, we present the details of our studies that lead us to the above conclusions. We first present the models we consider, and next, we discuss the static and then the dynamical properties of these models. Finally, we propose an experimental study that is well within the reach of current cold atom experiments and conclude the letter with the open threads initiated by this study.

\paragraph*{\emph{Models.}}

We first consider the paradigmatic SSH model subjected to an on-site quasiperiodic disorder(Model-I). The pure SSH model\cite{PhysRevLett.42.1698,PhysRevB.22.2099} can be described as a model of a single particle hopping in a one-dimensional lattice where the hopping amplitude is $t_1$ and $t_2$ between alternating pair of sites. The only scale in the pure SSH model is $\delta=t_2/t_1$. This model is known to exhibit a topological phase\cite{Asboth2016} when $\delta>1$. The Hamiltonian for Model-I is given by:
 \begin{equation}
     H_{SSH}=\sum_i t_1 c_i^\dagger c_{i+1}+t_2 c_{i+1}^\dagger c_{i+2} + \lambda \cos(2 \pi b i+
     \phi) c_i^\dagger c_{i} + h.c.
 \end{equation}
 The on-site potential is characterized by the
on-site modulation strength $\lambda$, an irrational period $1/b$, and the phase parameter $\phi$, which we set to zero without loss of any generality.

The second model we consider is the famous Rice-Mele(RM) model\cite{PhysRevLett.49.1455} subjected to an on-site quasi-periodic disorder(Model-II). The pure RM model is an extension of the SSH model with a staggered on-site potential and is also known to host a topological phase. The Hamiltonian for the Model-II is given by:
 \begin{equation}
     H_{RM}=\sum_i t_1 c_i^\dagger c_{i+1}+t_2 c_{i+1}^\dagger c_{i+2} + (-1)^i \lambda  \cos(2 \pi b i) c_i^\dagger c_{i} + h.c.
 \end{equation}
 Symbols $\lambda,b,\delta$ and $\phi$ stand for the same quantities as before(Model-I). 
  
The third model we consider is the Aubry-Andre model with an additional staggered potential of strength $\Delta$(Model-III). The Hamiltonian is given by:
 
 \begin{equation}
       H_{SP}=t\sum_i c_i^\dagger c_{i+1} + (\lambda +\Delta (-1)^i) \cos(2 \pi b i) c_i^\dagger c_{i} + h.c.
 \end{equation}
The fourth model we consider is a model of bi-chromatic potential( Model-IV). This is historically\cite{aa_multifrac} the first extended AA model, and this model is often studied in experiments. The Hamiltonian for Model-IV is given by, 
 \begin{equation}
       H_{BC}=t\sum_i c_i^\dagger c_{i+1} + \lambda  \cos(2 \pi b i) c_i^\dagger c_{i} +\Delta \cos(4 \pi b i) + h.c.
 \end{equation}
  
Notably, Models III and IV have two different periodic modulations both in the on-site energy scale, but for Models I and II, the two different periodic modulations are in the on-site and hopping energy scale, respectively. 

The salient feature that is common to all these models is the presence of two energy scales that are characterized by two different periodicities, which are different from the lattice periodicity. In the rest of this letter, we will study the interplay of these two scales.

\begin{figure}[t]
\centering
\subfigure[]{
\includegraphics[width=0.22\textwidth,height=3.0cm
]{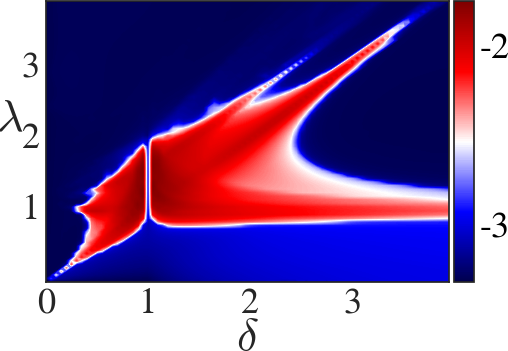}
\vspace*{-2.2em}}
 \subfigure[]{
\includegraphics[width=0.22\textwidth,height=3.0cm]{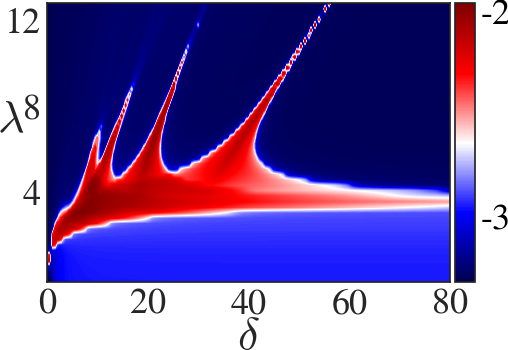}
 \vspace*{-2.2em}}
  \subfigure[]{
\includegraphics[width=0.22\textwidth,height=3.0cm]{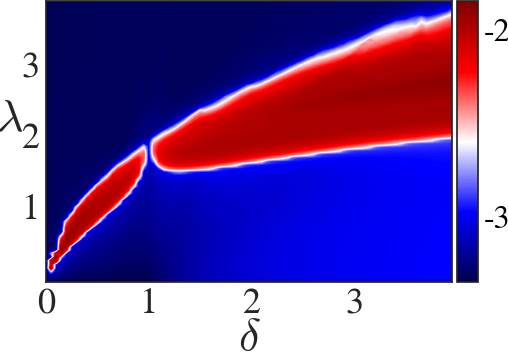}
 \vspace*{-2.2em}}
 \subfigure[]{
\includegraphics[width=0.22\textwidth,height=3.0cm]{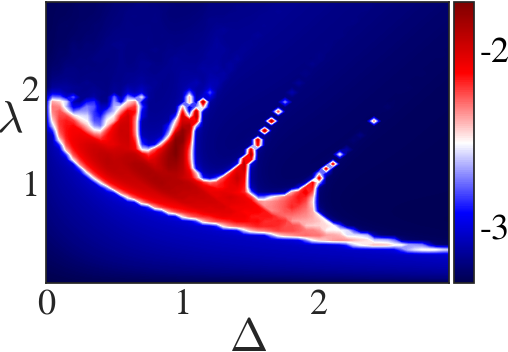}
 \vspace*{-2.2em}}
\caption{$\eta$ (represented in color code) phase diagrams for Models I-III with system size, $L=1974$. (a) Model I: $\eta$ plotted against $\lambda$ and $\delta$ with $b=\frac{\sqrt{5}+1}{2} +\frac{\sqrt{13}+3}{2}$(sum of golden ratio and bronze ratio), (b) Model I: $\eta$ plotted against $\lambda$ and $\delta$ with $b = \sqrt{2}+1$(silver ratio) (c) Model II: Plotted against $\lambda$ and $\delta$ with $b=\frac{\sqrt{5}+1}{2} +\frac{\sqrt{13}+3}{2}$, (d) Model III: $\eta$ plotted against $\lambda$ and $\Delta$ with $b =\sqrt{2}+1$.}
\label{fig:eta_pd1}
\vspace{-5mm}
\end{figure}  
\paragraph*{\emph{Statics.}}
We study the statics of single-particle states of models I-IV using the standard probes of localisation transition, namely the inverse participation ratio and the normalized participation ratio.

Inverse participation ratio(IPR) for $k$-th eigenstate of the Hamiltonian is given by $IPR(k)=\frac{\sum_k (\phi_i^k)^4}{(\sum_k (\phi_i^k)^2)^2}$, where $\phi_i^k$ denotes the $k$-th eigenstate  at site $i$. For a completely delocalised/extended state, IPR scales as 
$\sim 1/L$, where $L$ is the system size, for a completely localised state, IPR doesn't scale with size and remains finite. A power law form of scaling of IPR, $\sim \frac{1}{L^\gamma}$, where $0<\gamma<1$, implies critical nature of states. A complimentary quantity 
to study localisation transition is the normalized participation ratio, given by $NPR(k)=\frac{1}{L \sum_k (\phi_i^k)^4}$, which scales in the opposite way as IPR, by virtue of being inverse of IPR. 
For an individual state, if one takes a product of IPR and NPR, it will be unity due to this inverse relation, but as one computes the average of IPR and NPR over all the states and then takes the product as $\eta=\langle IPR \rangle \langle NPR \rangle$\cite{aa_me_bichrome_2,aa_me_bichrome_4}, the product $\eta$
remains finite for a large enough system only if there is a mobility edge in the spectrum.

In Figure \ref{fig:eta_pd1}, we show the $\eta$ phase diagram for Models I-III in the plane of two energy scales $\delta$(Model I and II)/$\Delta$(Model-III) and $\lambda$ for different quasi-periodic periodicities $1/b$.  Due to an interplay of the periodic hopping modulation scale $\delta$ and quasi-periodic potential amplitude $\lambda$,  the quasi-periodic SSH model shows a very interesting re-entrant quantum phase transition\cite{ssh_reentrant} i.e. for a given value of $\delta$ as one increases the $\lambda$, some of the single-particle eigenstates undergoes through a cascade of delocalisation-localisation-delocalisation-localisation transitions. In Figure \ref{fig:eta_pd1}a and \ref{fig:eta_pd1}b, we show that the energy scale and the number of re-entrant localisation transitions depend strongly on the choice of irrational number $b$ for the SSH model. Note that it is only the fractional part of $b$(denoted as $\{b\}$) that is responsible for quasi-periodic behaviour\footnote{Any two irrational numbers with fractional part $\{b\}$ and $1-\{b\}$ will have identical quasi-periodic behaviour due to the periodic properties of sinusoidal functions.}. In Figure \ref{fig:eta_pd1}a, we choose 
$b=\frac{\sqrt{5}+1}{2} +\frac{\sqrt{13}+3}{2}$ for which $\{b\}=0.920...$, and we see the re-entrant energy scale is in the range $\delta=2.5-3.5$ and $\lambda=1-3$. In Figure \ref{fig:eta_pd1}b, we choose $b=\sqrt{2}+1$ for which $\{b\}=0.414...$, and we see the multiple re-entrant localisation transitions in the ranges $\delta=15-55$ when  $\lambda$ is in the range $1-10$. This behaviour of the SSH model should be contrasted with the earlier studies\cite{ssh_reentrant} that focus only on the irrational number $b=\frac{\sqrt{5}+1}{2}$ for which $\{b\}=0.618...$ and demonstrated a single re-entrant transition at $\delta=20-30$, $\lambda=3-10$. 

\begin{figure}[t]
\centering
\subfigure[]{
\includegraphics[width=0.22\textwidth,height=3.0cm]{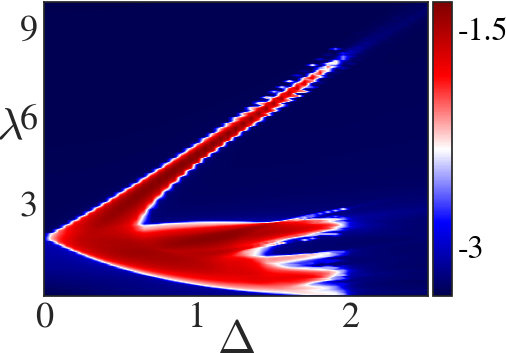}
 \vspace*{-2.2em}}
\subfigure[]{
\includegraphics[width=0.22\textwidth,height=3.0cm]{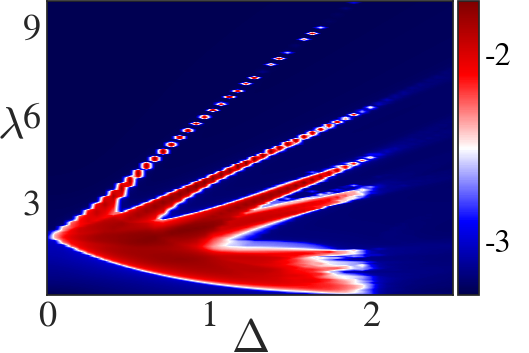}
\vspace*{-2.2em}}
\caption{$\eta$ (represented in color code) phase diagrams for Model IV with system size, $L = 1974$. (a) $\eta$ plotted against $\lambda$ and $\Delta$ with $b = \frac{\sqrt{13}+3}{2}$(the bronze ratio). (b) $\eta$ plotted against $\lambda$ and $\Delta$ with $b = \frac{\sqrt{29}+5}{2}$(nickel ratio).}
\label{fig:eta_pd2}
\vspace{-5mm}
\end{figure}

In Fig. \ref{fig:eta_pd1}c,  we plot the  $\eta$-phase diagrams of Model II(RM model), which shows a qualitatively similar re-entrant phase diagram as Model-I but the scale of the transition as a function of the irrational periodicity differs strongly between these models. In Fig. \ref{fig:eta_pd1}d, we show the  $\eta$-phase diagrams of model III; for a given value of $\lambda$ as one varies $\delta$, the model shows a cascade of re-entrant transitions\cite{2109.09621} where the number of such transitions is controlled by the choice of an irrational number. 

Next, we present our results for Model-IV, which is simply the nearest-neighbour tight-binding model with two quasi-periodic onsite potentials where the potentials have amplitude $\lambda$ and $\Delta$ with periodicities $1/b$ and $1/2b$ respectively. In Fig. \ref{fig:eta_pd2}a, we plot $\eta$-phase diagram in the plane of two energy scales $\delta$ and $\lambda$ for Model-IV for $b=\frac{3+\sqrt{13}}{2}$(bronze ratio, $\{b\}=0.302...$), and we find a cascade of re-entrant localisation  transitions in the range $\Delta=0.5-2$ and $\lambda=0-8$.In Fig.\ref{fig:eta_pd2}b we plot the $\eta$-phase diagram for Model-IV by choosing $b=\frac{5+\sqrt{29}}{2}$(nickel ratio, $\{b\}=0.192...$), and we find the number of re-entrant transitions increase. 
 
\begin{figure}[t]
\begin{center}
\includegraphics[width=0.49\columnwidth]{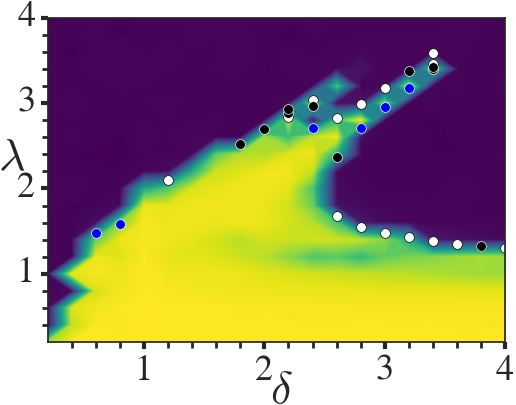}
\includegraphics[width=0.49\columnwidth]{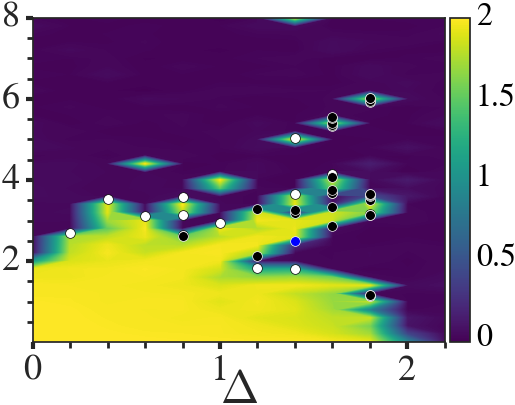}
\caption{Dynamical($\beta$, represented in color code) phase diagram for Model-I and -IV, respectively. Left: $\beta$ plotted for system size, $L = 1974$ with the choice of $b=\frac{\sqrt{5}+1}{2} +\frac{\sqrt{13}+3}{2}$(sum of golden ratio and bronze ratio), Right: $\beta$ plotted for system size, $L = 1974$ with the choice of $b = \frac{\sqrt{29}+5}{2}$(nickel ratio). In these figures we mark distinctly the boundary points for sub-diffusive(white), super-diffusive(black) and short time sub-diffusive but long time super-diffusive(blue) spread of an initially localised wave-packet.}
\label{fig:m2_pd1}
\end{center}
\vspace{-5mm}
\end{figure}

\emph{Notably, the frequency(or periodicity) dependence of the energy scale of DLT in any of these four models can't be connected to  the properties of its parent models i.e. the pure SSH/RM model or the pure AA model. Thus this frequency(or periodicity) dependent scale is an emergent phenomenon.}

A detailed analysis(in Supplementary Material) of the nature of individual eigenstates suggests that in the critical region, the spectrum is a mix of critical, localised and de-localised states. A thorough characterization of the nature of these critical states will be presented in a follow up work. In this letter, we present the dynamical properties at the critical parametric regions of Models I-IV. 

\paragraph*{\textbf{Dynamics}}
Understanding the temporal spread of a wave-packet under the quantum time evolution of a quasiperiodic hamiltonian is crucial for theoretical understanding\cite{wavepacket_1,wavepacket_2,sanyal_aa} of the critical point of DLT, and also one of the most important observable in the current state of the art cold atom experiments
\cite{aa_expt_1,aa_expt_2d,aa_expt_m2_1,aa_expt_m2_2,
aa_expt_slow_dynamics,aa_expt_subdiffusion_bosonic,aa_expt_spme,
aa_expt_spme1} in QP and disordered systems. However, very little is known about the dynamical signature of critical phases with mobility edge that has coexisting spectra of different kinds of states\cite{m2_gaa}. 

In this letter, we further study the temporal spread of a wave-packet under the quantum time evolution of Models I-IV using two different initial wave-packets. 
In the first case, we consider the spread of a wave-packet $\psi_i(t)$ which is initially localised at the central site of the lattice of size $N$, i.e. $\psi_i(0)=\delta_{i,N/2}$. We compute the time evolution of this initial wave packet by exact diagonalization and plot the width of this wave packet($\sigma(t)$) as a function of time, given by $\sigma(t)=\sum_i (i-\bar{i})^2|\psi_i(t)|^2$, where $\bar{i}$ is the position expectation value of the wavepacket. Typically, $\sigma(t)$ scales as a power law as $\sigma(t)\sim t^\beta$, where the dynamical exponent $\beta$ characterizes the dynamics. For systems with ballistic dynamics, $\beta=2$; for systems with diffusive dynamics $\beta=1$ and for systems which are localised or sub-power law dynamics, $\beta \sim 0$. 
On the other hand, $1<\beta<2$ indicates super-diffusive dynamics, and $0<\beta<1$ indicates sub-diffusive dynamics. 

\begin{figure}[t]
\begin{center}
\includegraphics[width=\columnwidth]{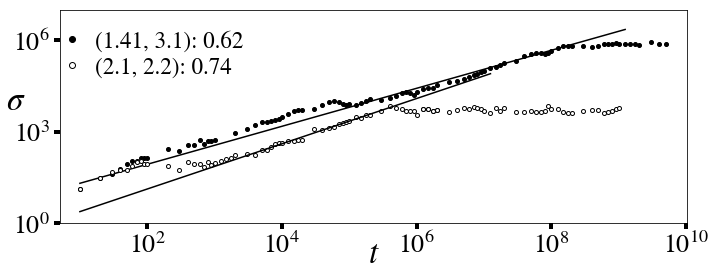}
\includegraphics[width=\columnwidth]{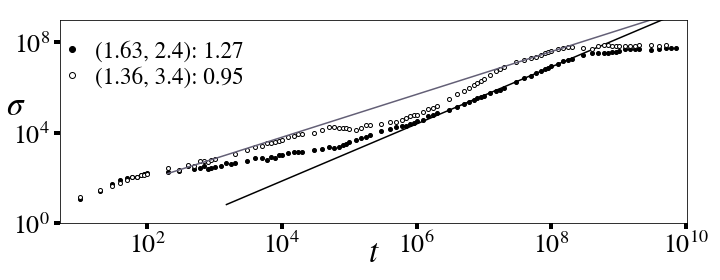}
\caption{$\sigma$ vs $t$ plots of an initially localised wave-packet for representative examples from Model-II, where $b=\frac{\sqrt{5}+1}{2}$ and $L=150050$. In the legends, we display $(\lambda,\delta):\beta$, where the $\beta$ is extracted by fitting the long-time behaviour. 
Top: At $(\lambda,\delta)$ value $(1.41,3.1)$, we find a sub-diffusive spread for upto $t \sim 10^8$, till the wave-packet hits the boundary. At $(2.1,2.2)$, we find a sub-diffusive spread till $t \sim 10^6$ that gets localised well before hitting the boundary. Bottom: At $(1.62,2.4)$, we find a sub-diffusive spread till $t \sim 10^5$, that becomes super-diffusive at time range $10^6-10^8$, till the wave-packet hits the boundary. At $(1.36,3.4)$, we see a diffusive spread till $t \sim  10^5$ that becomes super-diffusive in time $10^5-10^7$, and eventually diffusive again at time $10^7-10^8$, till the wave-packet hits the boundary.}
\label{fig:m2_t}
\end{center}
 \vspace{-2em}
\end{figure}
In Fig. \ref{fig:m2_pd1}, we plot the dynamical($\beta$) phase diagram for Model-I and IV in the plane of the two relevant energy scales $\delta$(or $\Delta$) and $\lambda$, for QP periodicity $b=\frac{\sqrt{5}+1}{2}$ and $L=1974$. In the supplementary material, we plot the $\beta$ phase diagram for Model-II and III at $L=1974$, and we show that the phase diagram remains qualitatively the same at a larger system size of $L=13530$(we plot only Model-II). The dynamical($\beta$) phase diagram (Fig. \ref{fig:m2_pd1}) shows a delocalised, critical, and localised phase. Notably, even if there is a critical region as per the $\eta$-phase diagram(Fig. \ref{fig:eta_pd1}) where the critical state can coexist with localised and or delocalised states, critical dynamics is visible only when the spectrum is not dominated by delocalised states. As an example, consider the re-entrant transition in Fig. \ref{fig:m2_pd1}a and in Fig. \ref{fig:eta_pd1}a for $\delta=2.4$; when $\lambda=0-1.45$, all the states in the spectrum are delocalised and thus the dynamics shows complete delocalisation.  $\lambda=1.45$  marks the onset of DLT where some of the states become critical and eventually the spectrum is a mix of critical and localised states. At the onset of the point when the spectrum is a mix of critical and localised states(see Supplementary Material), one expects to see a critical signature in the long time evolution of an initially localised wave-packet, as we see for  $\lambda=1.45, \delta=2.4 $ in Fig. \ref{fig:m2_t}.

We find a remarkably rich spectrum of critical dynamics at the phase boundary that includes sub-diffusive, super-diffusive and diffusive spread of wave-packet, and also two different kinds of spreads at a short and long time. In Fig. \ref{fig:m2_pd1} we mark the phase boundary points for each of these regimes at system size $L=1974$.

 In Fig. \ref{fig:m2_t}, we plot $\sigma$ vs $t$ at a much larger system size of $L=150050$ for representative cases of each kind of behaviour in Model-II, where we choose $b=\frac{\sqrt{5}+1}{2}$.  At $(\lambda,\delta)$ value $(1.41,3.1)$, we find a sub-diffusive spread up to a very long time, $t \sim 10^8$, till the wave-packet hits the boundary. At $(2.1,2.2)$, we find a sub-diffusive spread till $t \sim 10^6$, the wave-packet gets localised after that time and never hits the boundary. At $(1.62,2.4)$, we find a sub-diffusive spread till $t \sim 10^5$, after that the spread becomes super-diffusive at time range $10^6-10^8$, till the wave-packet hits the boundary. At $(1.36,3.4)$, we see a diffusive spread till $t \sim  10^5$ that becomes super-diffusive in time $10^5-10^7$, and eventually diffusive again at time $10^7-10^8$, till the wave-packet hits the boundary. To the best of our knowledge, this is the first example of a QP system to demonstrate such a rich range of dynamical behaviour, while we show only results from Model-II here, all other models studied here also have similarly rich dynamical characteristics.

Our results generate an intriguing question about the dynamical characteristics of disordered systems, that we pose as following: in the long time dynamics of an initially localised wave-packet one expects a contribution from all the eigenstates with a phase factor determined by the eigenvalues, what properties of mixed spectra(in localisation characteristics) of eigenstates controls the short and long time dynamical behaviour of an initially localised wave packet? \footnote{we present a preliminary analysis in the supplementary material, a detailed account of this problem will be presented elsewhere}. A thorough approach to answer this question will be important to understand the dynamical characteristics of a large class disordered system with mobility edge, and our study shows that the "two-frequency" class of QP models is an excellent playground for the same. 

\begin{figure}[b]
\begin{center}
\includegraphics[width=0.49\columnwidth]{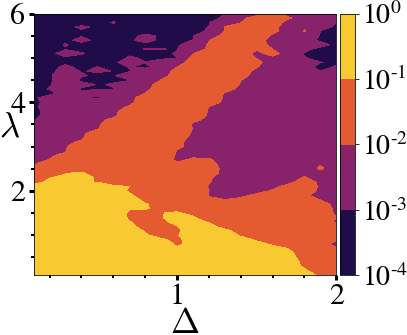}
\includegraphics[width=0.49\columnwidth]{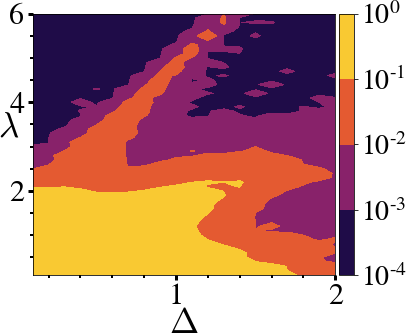}
600\caption{$\mathcal{D}$ (represented in color code) phase diagrams for Model-IV averaged over 10 phases, plotted against $\lambda$ and $\Delta$ for system size, $L = 600$(Left) and $L=650$ (Right). Left: $b = \frac{\sqrt{5}+1}{2}$ approximated by ${b} = \frac{377}{610}$, Right: $b = \frac{\sqrt{13}+3}{2}$(bronze ratio) approximated by ${b} = \frac{201}{664}$.}

\label{fig:edge_density}
\end{center}
\vspace{-5mm}
\end{figure}

Next we discuss the second case of initial wave-packet, where the wave-packet is initially localised at the left half of the lattice, we calculate the probability of finding the particle in the rest of the lattice as a function of time. This quantity is dubbed as edge density ($\mathcal{D}$)\cite{aa_me_bichrome_2}, defined as $\mathcal{D}=\sum_{i \in Right} |\psi_i(t)|^2$, where $\psi_i(0)=\frac{1}{\sqrt{L/2}} 
~~\forall i \in$ Left, otherwise, $\psi_i(0)=0 $. In Fig. \ref{fig:edge_density}a we plot $\mathcal{D}$(color coded) in the $\lambda-\Delta$ plane for Model-IV at $b=\frac{\sqrt{5}-1}{2}$(Left) and $b=\frac{\sqrt{13}+3}{2}$(Right). We approximate these two irrational numbers as a ratio of coprime integers given by $377/610$ and $201/664$, respectively. Notably, in experiments the quasi-periodic potential is generated by superposing two optical lattices of co-prime periodicity, in Fig. \ref{fig:edge_density}, we demonstrate that a clear signature of re-entrant phase transition can be seen in such a set-up in the quantity, $\mathcal{D}$, that is considered a more direct measure of the presence of non-localised states. The energy scale and number of transitions can be tuned by choosing the optical lattice periodicities.

\paragraph*{\textbf{Experiments}} In a recent experiment\cite{aa_expt_spme} bichromatic QP model(Model-IV) is realised by superimposing a primary optical lattice of wavelength $532.2$ nm, and two deep orthogonal lattices of wavelength $738.2$ nm. Furthermore the experiment studied the expansion of an initial state of non-interacting fermionic  $^{40}K $ atoms loaded in the middle third of the lattice initially, the expansion of which is captured by the edge-density $\mathcal{D}$. All the parameters of Model-IV can be tuned in this experimental set-up, $\lambda$ and $\Delta$ can be controlled by the depth of the primary and detuning optical lattice and a different $\{b\}$ can be chosen by changing the primary and detuning lattice wavelengths. Thus our first two main results regarding the controllability of number and energy scale of DLT can be verified within an existing experimental set up. 

Our third main result regarding an wide range of dynamical behaviour can be also verified within the existing state of the art cold atom experiments that routinely study\cite{aa_me_bichrome_2,aa_expt_subdiffusion_bosonic,aa_expt_m2_1,aa_expt_m2_2} spatial expansion of an initially compact gas of gold atoms in a quasiperiodic lattice. A recent experiment \cite{photonic_ssh} on photonic quasicrystals demonstrating re-entrant DLT is also a promising direction for the experimental realization of our main results.

\paragraph*{\textbf{Conclusion}}

The main results presented in this letter are twofold. First: we discover a new "knob" to tune the energy scale of delocalisation-localisation transition and number of re-entrant transitions in a class of QP models that has two periodic scales with periodicities different from the lattice periodicity. Second: we found rich dynamical characteristics of the critical dynamics in these models that includes the trifecta of sub-diffusive, diffusive and super-diffusive temporal spread of an initially localised wave-packet in a single QP model. Moreover, we found time-dependent change in the dynamical characteristics in these systems, e.g. an initially localised wave-packet spreads sub-diffusively at the short time but eventually crosses over to spread super-diffusively in the long time limit, and so on. Notably this is the first example(to the best of our knowledge) of super-diffusive dynamics in the single particle physics of QP systems\footnote{The only other example of super-diffusive dynamics is quasi-periodic systems\cite{barlev_superdiffusion} require interaction and speculated to be transient.}. We showed both of these two results are well within the reach of current state of the art cold atom and photonic experiments but perhaps more importantly these results opens several threads to be explored, we conclude our letter with a brief outline of the same.

The phenomenology of an wider range of models with re-entrant localisation transition with two periodic scales and their rich spectral properties in dynamical behaviour begs a deeper mathematical understanding. It could be an worthwhile undertaking to extend Avila's  "Global theory of one-frequency Schrödinger operators"\cite{avila_global} towards two-frequency Schrödinger operators, understanding the spectral properties of metal-insulator transitions\cite{aa_math_1,ten_martini} in such systems might be of great value towards understanding the sub-to-super diffusive dynamical behaviour we observe.  

QP models holds an important position in physics at the crossroads of many-body localisation phenomena\cite{iyer_mbl,modak_me,dassarma_me}, out-of-equilibrium physics\cite{sanyal_aa}, physics of non-hermitian quantum mechanics\cite{non-hermitian-AA}, open quantum systems\cite{sanyal_aa}, and topological phases\cite{aa_topology_1,aa_topology_2}. Our finding of a new "knob" and the rich dynamics in certain QP systems  might open new dimensions of understanding and explorations in these areas.

\begin{acknowledgments}
\paragraph*{\textbf{Acknowledgments}}
S.~S. acknowledges support from Science and Engineering Research Board (Department of Science and Technology) Govt. of India, under grant no. SRG/2020/001525 and an internal start up grant from Indian Institute of Science Education and Research, Tirupati. D.~J. acknowledges support from Science and Engineering Research Board (Department of Science and Technology) Govt. of India through the NPDF fellowship.
\end{acknowledgments}

\bibliography{main}

\newpage
\onecolumngrid
\appendix
\newpage

\section{Supplementary Section}

In this Supplemental Material, we present additional evidence which support the key findings described in the main text. We first present the results of the dynamics properties and then the statics properties. 

\section{Dynamics}

\begin{figure}[b]
\begin{center}
\includegraphics[width=7.55cm]{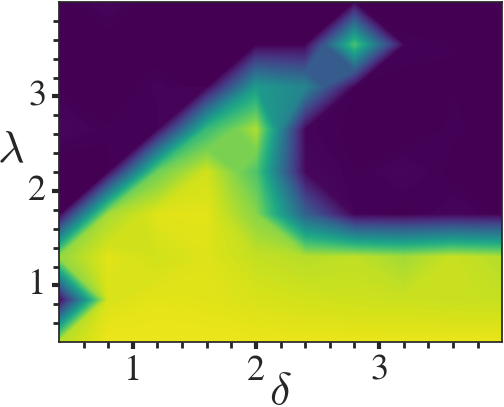}
\includegraphics[width=8cm]{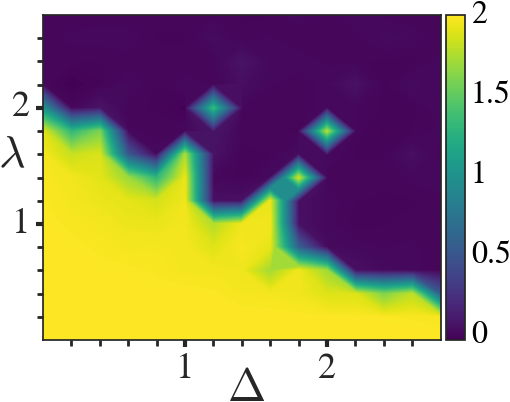}
\caption{Dynamical($\beta$) phase diagram for Model-II \& -III for system size $L = 1974$ and choice of $b = \frac{\sqrt{5}+1}{2}$.}
\label{fig:m2_pd_sup1}
\end{center}
\end{figure}

\begin{figure}
\begin{center}
\includegraphics[width=8cm]{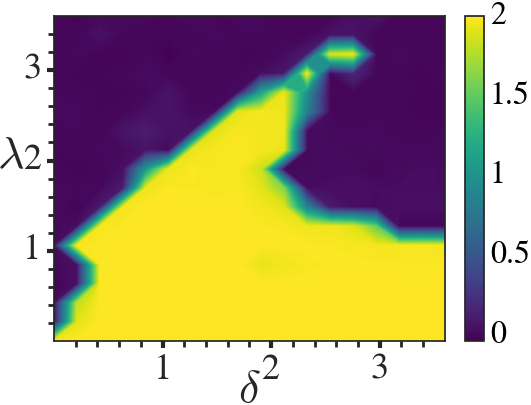}
\caption{Dynamical($\beta$) phase diagram for Model-II for system size, $L = 13530$ and choice of $b = \frac{\sqrt{5}+1}{2}$.}
\label{fig:m2_pd_sup2}
\end{center}
\end{figure}

In Fig. \ref{fig:m2_pd_sup1} we plot the dynamical($\beta$) phase diagram for Model-II and III in the plane of the two relevant energy scales $\delta$(or $\Delta$) and $\lambda$, for QP periodicity $b=\frac{\sqrt{5}+1}{2}$. This plot is to supplement analogous plots on Models-I and IV in the main text(Fig. \ref{fig:m2_pd1}). In Fig. \ref{fig:m2_pd_sup2} we show the dynamical($\beta$) phase diagram for Model-II for a much larger system size, the phase diagram remained qualitatively same except the boundary between different phases are much more clearly visible at a larger size. 

\section{Statics}
In this section we plot the state resolved IPR for some of the representative points from Model-II that exhibits critical dynamics and thus expected to have critical states in the spectrum. 
\begin{figure}
\begin{center}
\includegraphics[width=8cm]{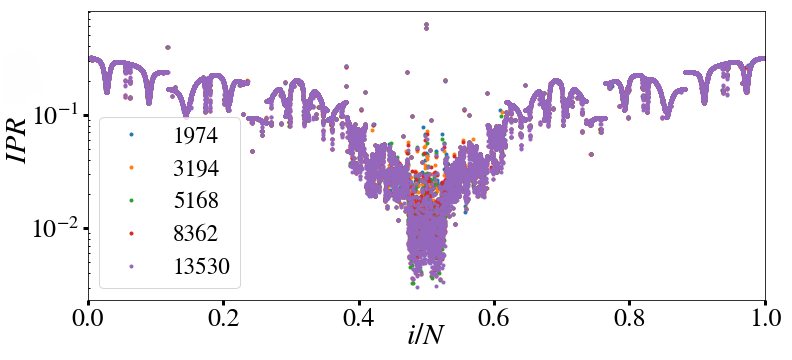}
\includegraphics[width=8cm]{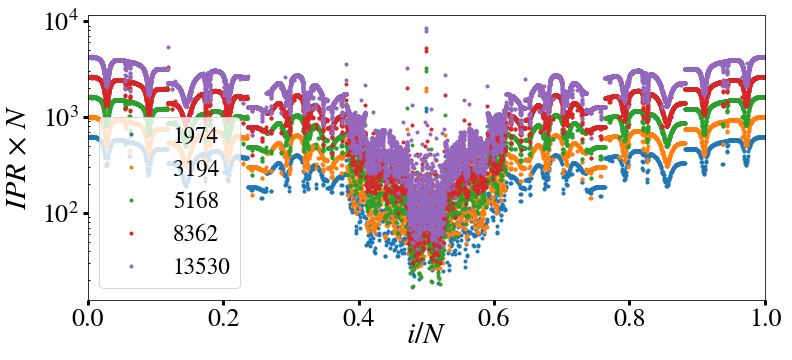}
\caption{State resolved IPR of the point $\lambda = 1.41, \delta = 3.1$ in Model-II with $b = \frac{\sqrt{5}+1}{2}$.The point exhibits sub-diffusive dynamics at long time.}
\label{fig:state_ipr_sub}
\end{center}
\end{figure}
In Fig. \ref{fig:state_ipr_sub} we plot the state resolved IPR for $\lambda = 1.41, \delta = 3.1$ in Model-II with $b = \frac{\sqrt{5}+1}{2}$. For this example an initially localised wave-packet spreads sub-diffusively at very long time as shown in Fig. \ref{fig:m2_t}. In Fig. \ref{fig:state_ipr_sub}a we show that almost all the states are localised as the state resolved IPR for different size collapse on each other except for a few states near the centre of the spectrum. Fig. \ref{fig:state_ipr_sub}b we plot the same state resolved IPR times the system size to demonstrate that those few states are not delocalised either, so these are critical states.
\begin{figure}
\begin{center}
\includegraphics[width=8cm]{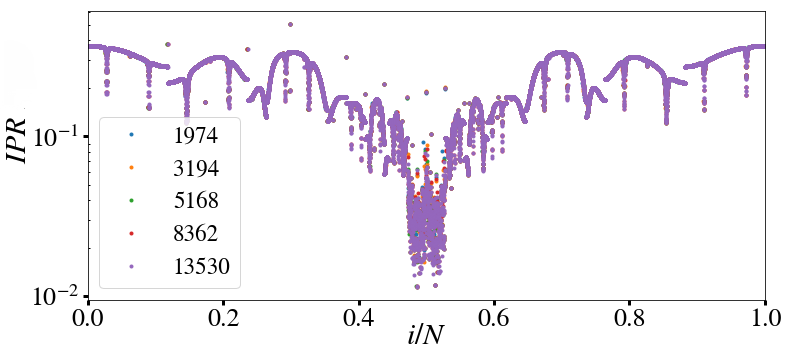}
\includegraphics[width=8cm]{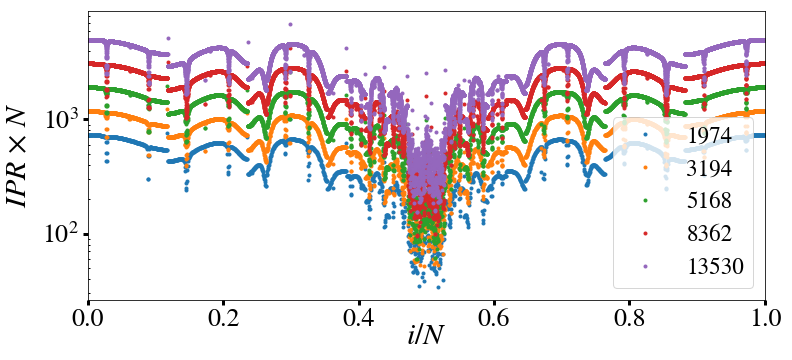}
\caption{State resolved IPR of the point $\lambda = 2.1, \delta = 2.2$ in Model-II with $b = \frac{\sqrt{5}+1}{2}$. This point exhibits sub-diffusive dynamics at short time and gets localised at long time.}
\label{fig:state_ipr_sub_loc}
\end{center}
\end{figure}
In Fig. \ref{fig:state_ipr_sub_loc}a we plot the state resolved IPR for $\lambda = 2.1, \delta = 2.2$ in Model-II with $b = \frac{\sqrt{5}+1}{2}$. For this case an initially localised wave-packet will spread sub-diffusively at short time but eventually it will get localised, as shown in Fig. \ref{fig:m2_t}. Notably the state resolved IPR for this example in Fig. \ref{fig:state_ipr_sub_loc} is qualitatively very similar to that of a sub-diffusive dynamics case in Fig. \ref{fig:state_ipr_sub}. 
\begin{figure}[ht]
\begin{center}
\includegraphics[width=8cm]{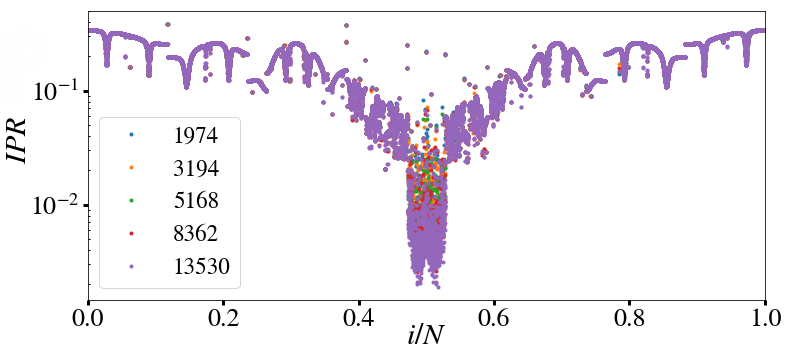}
\includegraphics[width=8cm]{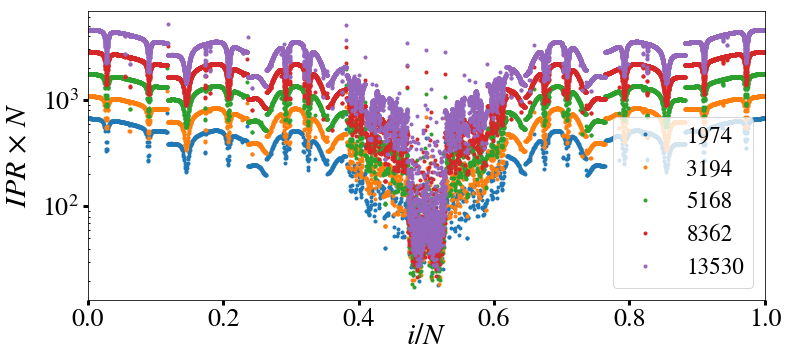}
\caption{State resolved IPR of the point $\lambda = 1.63, \delta = 2.4$. in Model-II with $b = \frac{\sqrt{5}+1}{2}$. This point exhibits Superdiffusive dynamics at long time.}
\label{fig:state_ipr_sup}
\end{center}
\end{figure}
\begin{figure}[ht]
\begin{center}
\includegraphics[width=8cm]{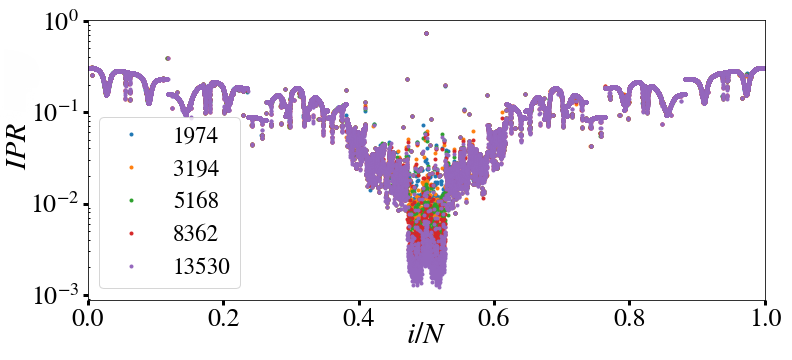}
\includegraphics[width=8cm]{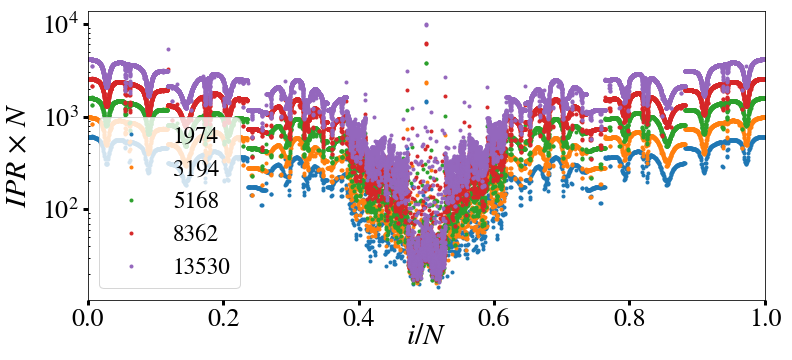}
\caption{State resolved IPR of the point $\lambda = 1.36, \delta = 3.4$. in Model-II with $b = \frac{\sqrt{5}+1}{2}$. This point exhibits diffusive dynamics at long time.}
\label{fig:state_ipr_diff}
\end{center}
\end{figure}
In Fig.\ref{fig:state_ipr_sup} we plot the state resolved IPR for $\lambda = 1.63, \delta = 2.4$ in Model-II with $b = \frac{\sqrt{5}+1}{2}$. For this example an initially localised wave-packet spreads superdiffusively at very long time as shown in Fig. \ref{fig:m2_t}. In Fig. \ref{fig:state_ipr_sup}a we show that almost all the states are localised as the state resolved IPR for different size collapse on each other except for a few states near the middle of the spectrum. Fig. \ref{fig:state_ipr_sup}b we plot the same state resolved IPR times the system size to demonstrate that those few states are not delocalised, thus these states are critical states. 

In Fig.\ref{fig:state_ipr_diff} we plot the state resolved IPR for $\lambda = 1.36, \delta = 3.4$ in Model-II with $b = \frac{\sqrt{5}+1}{2}$. In this case an initially localised wave packet will spread almost diffusively at long time as shown in Fig. \ref{fig:m2_t}. Notably here the critical state appears only near the centre of the spectrum, this can be contrasted with the critical point of Aubry-Andre model where an initially localised wave-packet spreads diffusively but all the states of the spectrum are critical. 

Understanding the critical nature of each of the wave-functions is important to understand the long time dynamics of an initially localised wavepacket, as it is the wave-function characteristics and eigenvalues that entirely determine the dynamics. In this supplementary material we demonstrated the nature of the wave-function through state resolved IPR. Here we highlight some of the visible qualitative differences in the spectrum that leads to a significant difference in the long-time behaviour of the wave-packet spreading, but we also observe that it might be important to analyse the subtle features like the fraction of critical states and  
the nature of criticality (a more detailed account of these aspects will be presented elsewhere).  
\end{document}